\documentclass[twocolumn,times]{aastex631}
\hypersetup{linkcolor=green,citecolor=magenta,filecolor=cyan,urlcolor=red}
\usepackage{rotating}
\usepackage{mathrsfs}
\usepackage{wrapfig}
\usepackage{hyperref}
\usepackage{graphicx}

\usepackage{xcolor}

\begin{document}

\title{A Deep VLA Search for a Persistent Radio Counterpart to the One‑off FRB~20250316A}

\author[0000-0003-4341-0029]{Tao An}
\affiliation{Shanghai Astronomical Observatory, Chinese Academy of Sciences, 80 Nandan Road, Shanghai 200030, China}
\affiliation{Key Laboratory of Radio Astronomy and Technology, Chinese Academy of Sciences, A20 Datun Road, Beijing, 100101, P. R. China}

\author[0000-0002-7351-5801]{Ailing Wang}
\affiliation{Key Laboratory of Particle Astrophysics, Institute of High Energy Physics, Chinese Academy of Sciences, Beijing 100049, China}
\affiliation{Shanghai Astronomical Observatory, Chinese Academy of Sciences, 80 Nandan Road, Shanghai 200030, China}

\author[0009-0009-7749-8998]{Yu-Chen Huang}
\affiliation{Department of Astronomy, University of Science and Technology of China, Hefei 230026, China}
\affiliation{School of Astronomy and Space Science, University of Science and Technology of China, Hefei 230026, China}

\author[0009-0008-6783-5762]{Jia-Pei Feng}
\affiliation{Department of Astronomy, University of Science and Technology of China, Hefei 230026, China}
\affiliation{School of Astronomy and Space Science, University of Science and Technology of China, Hefei 230026, China}

\author{Yuanqi Liu}
\affiliation{Shanghai Astronomical Observatory, Chinese Academy of Sciences, 80 Nandan Road, Shanghai 200030, China}

\author[0000-0002-8366-3373]{Zhongli Zhang}
\affiliation{Shanghai Astronomical Observatory, Chinese Academy of Sciences, 80 Nandan Road, Shanghai 200030, China}
\affiliation{Key Laboratory of Radio Astronomy and Technology, Chinese Academy of Sciences, A20 Datun Road, Beijing, 100101, P. R. China}

\author[0000-0002-7835-8585]{Zi-Gao Dai}
\affiliation{Department of Astronomy, University of Science and Technology of China, Hefei 230026, China}
\affiliation{School of Astronomy and Space Science, University of Science and Technology of China, Hefei 230026, China}
\correspondingauthor{Tao An}
\email{antao@shao.ac.cn}

\begin{abstract}
Fast Radio Burst (FRB) 20250316A, detected by CHIME on 2025 March 16 with a fluence of $1.7\pm0.1$~kJy~ms and a dispersion measure of \(161.3\pm0.4~\mathrm{pc\,cm^{-3}}\), ranks among the brightest extragalactic FRBs at \(\sim 40\) Mpc. We obtained deep Karl G. Jansky Very Large Array follow-up at 15~GHz on 2025 April 5 and 9 and find no persistent radio source (PRS). Our best image reaches an rms of \(2.8~\mu\mathrm{Jy\,beam^{-1}}\), yielding a \(3\sigma\) upper limit of \(<8.4~\mu\mathrm{Jy}\) at the FRB position, corresponding to \(\nu L_\nu < 2.4\times10^{35}~\mathrm{erg\,s^{-1}}\). These results represent among the most stringent constraints for a non-repeating FRB, lying \(\gtrsim 3\) orders of magnitude below the \(\nu L_\nu\) of compact persistent radio sources around well-studied repeaters, thereby disfavoring bright magnetar-nebula scenarios and pointing to low-density, weakly magnetized environments. Interpreting our limit through pulsar-/magnetar-wind synchrotron frameworks places joint constraints on ambient density and engine power. If the empirical PRS--rotation-measure trend reported for repeaters extends to one-off sources, our limit implies $\vert \mathrm{RM} \vert \lesssim 30~\mathrm{rad\,m^{-2}}$, consistent with a clean magneto-ionic sight line and progenitor channels such as neutron-star mergers or giant flares from older magnetars.
\end{abstract}



\section{Introduction} \label{sec:intro}

Fast radio bursts (FRBs) are millisecond-duration, highly luminous radio transients of extragalactic origin, first discovered by \citet{Lorimer2007}. These enigmatic events release energies up to $\sim10^{40}$ erg in coherent radio emission, with peak flux density ranging from tens of mJy to up to kJy \citep{Petroff2019, Cordes2019, CHIME2021, Bruni2024, Bruni2025}. Over the past decade, more than 800 FRBs have been cataloged through wide-field surveys like the Canadian Hydrogen Intensity Mapping Experiment \citep[CHIME,][]{CHIME2021} and the Australian Square Kilometre Array Pathfinder \citep[ASKAP,][]{Shannon2018}, with recent estimates suggesting $\sim5-10\%$ identified as repeaters \citep{CHIME2021, Xu2023}. Repeating FRBs exhibit diverse burst morphologies, including periodicity in some cases \citep[e.g.][]{CHIME2020}, while non-repeating (or one-off) FRBs dominate the observed sample.

The progenitors of FRBs remain a central puzzle in high-energy astrophysics. Leading models invoke young magnetars \citep{Murase2016, Metzger2019, Margalit2019} or binary systems each of which contains at least one neutron star \citep{Wang2016, dai2016, Zhang2020, Wang2022}. The association of FRB 20200428A with the Galactic magnetar SGR 1935+2154 provided direct evidence for magnetar origins in at least some cases \citep[e.g.,][]{Bochenek2020, CHIME2020}, though extragalactic FRBs probe a broader parameter space up to redshifts $z \sim 1$ \citep{Macquart2020}.

A key breakthrough has been the identification of persistent radio sources (PRSs) co-located with some repeating FRBs. The first association was with FRB 121102A, where a compact ($\sim0.7$ pc), luminous ($\sim10^{39}$ erg s$^{-1}$) PRS was detected \citep{Chatterjee2017, Marcote2017}. Subsequent discoveries include PRSs linked to FRB 20190520B \citep{Niu2022}, FRB 20201124A\footnote{We note that the persistent radio emission associated with FRB 20201124A remains a candidate PRS, given its physical offset from the FRB localization and the relatively shallow constraint on its size, which is not comparable to those of confirmed PRSs \citep{Ravi2022, Piro2021, Nimmo2022, Bruni2024, Rahaman2025}. } \citep{Ravi2022}, and FRB 20240114A \citep{Bhusare2024, Bruni2025}. These non-thermal, synchrotron-emitting nebulae are potentially powered by magnetar winds or supernova remnants \citep{Margalit2018, Bruni2024}. To date, three confirmed PRSs are all associated with repeating FRBs in low-metallicity host galaxies \citep[e.g.,][]{Heintz2020}, hinting at a connection to young, massive stellar populations. However, no PRS has been firmly linked to a one-off FRB, leaving open whether all FRBs share similar progenitors \citep{Ravi2019}. 
% Non-detections in surveys of one-off FRBs suggest any associated PRSs must be fainter than $\sim10$ $\mu$Jy at GHz frequencies \citep{Law2022}. Deep radio continuum searches for PRSs are essential for constraining FRB models, as upper limits can rule out certain nebula sizes or luminosities \citep{Aggarwal2021}.
Deep imaging of localized, apparently non-repeating FRBs has yielded no compact persistent emission down to $\sim10$–$20~\mu$Jy at GHz frequencies \citep[e.g.,][]{Aggarwal2021}, implying any associated PRSs are substantially fainter than those seen near some repeaters \citep[e.g.,][]{Law2022}. Such deep radio-continuum searches and the resulting upper limits provide strong constraints on FRB environments by excluding regions of the allowed nebular size–luminosity parameter space. 

FRB 20250316A, discovered on 2025 March 16 by CHIME/FRB, stands out as one of the brightest extragalactic FRBs detected to date with an extraordinary fluence of $1.7 \pm 0.1$ kJy ms and ${\rm DM} = 161.3 \pm 0.4$ pc cm$^{-3}$ \citep{ATel17081}, making it a prime target for PRS searches. Initially associated with a nearby X-ray transient EP J120944.2+585060 \citep{ATel17100}, subsequent high-precision localizations established they are spatially independent phenomena \citep{ATel17086, ATel17119, GCN39886}. Using the full CHIME Outriggers array, \citet{CHIME20250316A} achieved unprecedented $68 \, \text{mas} \times 57 \, \text{mas}$ localization precision, enabling detailed study of the local environment at 13-parsec precision.
The FRB is localized within the nearby SBc spiral galaxy NGC 4141 at $\sim$40 Mpc ($z \approx 0.009$) \citep{ATel17114}.  The FRB location is offset by $190 \pm 20$ pc from the center of the nearest star-forming region, with local gas-phase metallicity of $\log (Z/Z_\odot) \sim -0.1$. 
The host galaxy of FRB~250316A has a modest star-formation rate of $\sim0.6~M_\odot \,\mathrm{yr}^{-1}$ \citep{CHIME20250316A}, which falls within the range of SFRs measured for the low-mass, actively star-forming hosts of canonical PRS repeaters (e.g., FRB20121102A and FRB20190520B) \citep{Heintz2020}. 
% The host galaxy has a modest star formation rate $\sim 0.6 \, M_\odot$ yr$^{-1}$ \citep{Kewley2005}, in contrast with the dwarf, actively star-forming hosts of repeating FRBs with detected PRSs, which typically have stellar masses $< 10^9 M_\odot$  and specific star formation rates $> 10^{-9} \, \text{yr}^{-1}$ \citep{Heintz2020}.

No repeat bursts were reported in continued CHIME monitoring \citep{ATel17081}, consistent with the non-detections by follow-ups with the Five-hundred-meter Aperture Spherical radio Telescope (FAST) and a European network of radio telescopes as part of the HyperFlash project, establishing a 15.3 mJy~ms fluence upper limit ($7\sigma$) at L band  \citep{LiY2025, ATel17124} and confirming FRB 20250316A as either genuinely one-off or having an extremely low repetition rate \citep{Li2025}. Its one-off nature, high brightness, and proximity enable unprecedented constraints on progenitor models.

In this Letter, we report Karl G. Jansky Very Large Array (VLA) observations of FRB 20250316A aimed at detecting or firmly constraining any PRS. 
We adopt a cosmology of $H_0 = 70$ km s$^{-1}$ Mpc$^{-1}$, $\Omega_m = 0.3$, $\Omega_\Lambda = 0.7$.

\section{Observations and Data Reduction}

Building upon the precise very long baseline interferometry (VLBI) localization  \citep{CHIME20250316A}, we conducted VLA observations (code: 25A-435, PI: T. An) of FRB 20250316A in D-array configuration over two epochs: 2025 April 5 (C-band: 4--8 GHz; X-band: 8--12 GHz; Ku-band: 12--18 GHz) and 2025 April 9 (Ku-band only). We selected Ku-band for both epochs due to its balance of resolution and sensitivity to compact synchrotron sources, with the April 9 focus on Ku band alone to prioritize resolution and imaging depth. Each epoch used 27 antennas with full polarization products, with observational parameters summarized in Table \ref{tab:obs}.

Data were processed using CASA following standard VLA calibration pipelines. We used 3C~286 for flux density and bandpass calibration. J1219+4829 (C- and X-band) and J1152+4939 (Ku-band) were selected as the gain calibrators for its proximity and stability. One iteration of phase-only self-calibration on the target field improved image dynamic range by $\sim$20\%. Target field visibilities were imaged using the \texttt{tclean} task with Briggs weighting (robust = 0.5) to optimize point-source sensitivity. Multi-scale \texttt{clean} algorithm with scales of 0, 8, and 20 times the pixel size recovered potential extended emission. Detailed calibration procedures and imaging strategies are provided in Appendix \ref{app:reduction}.

\section{Results}

\begin{table}
\centering
\caption{VLA Observational Parameters}
\label{tab:obs}
\begin{tabular}{lcc}
\hline
Parameter & April 5 & April 9 \\
\hline
Bands & C, X, Ku & Ku \\
Central frequencies (GHz) & 6, 10, 15 & 15 \\
Bandwidth (GHz) & 4, 4, 6 & 6 \\
total time (hours) & 1.8 & 1.8 \\
Beam size (arcsec) & 14$\times$9, 9$\times$6, 7$\times$4 & 6$\times$4 \\
rms noise ($\mu$Jy) & 5.3, 5.7, 5.9 & 2.8 \\
$3\sigma$ upper limit ($\mu$Jy) & 15.9, 17.1, 17.7 & 8.4 \\
\hline
\end{tabular}
\end{table}

No significant persistent radio counterpart was detected at the FRB position across all observed bands and epochs. Table \ref{tab:obs} summarizes the achieved noise levels and corresponding 3$\sigma$ upper limits. The observed root-mean-square (rms) levels of $\sim$3--7 $\mu$Jy beam$^{-1}$ agree well with theoretical expectations of $\sim$3--5 $\mu$Jy beam$^{-1}$ indicating no significant systematic anomalies (detailed noise analysis in Appendix \ref{app:reduction}). We searched for sources within a 5$''$ radius around the FRB position, justified by CHIME localization uncertainty ($\sim$3$''$) and beam FWHM ($\sim$3$''$ at 15 GHz, scaling to $\sim$8$''$ at 6 GHz for conservative coverage). No statistically significant ($>3\sigma$) radio source was detected in any Stokes parameter. We adopt a $3\sigma$ threshold consistent with previous PRS searches (e.g., \citealt{Law2022,Aggarwal2021}). 
% At our achieved noise levels, this corresponds to a $< 0.3\%$ probability of spurious detection.
At our achieved noise levels, this corresponds to a $< 0.4\%$ probability of spurious detection, calculated assuming Gaussian noise statistics where the chance of a random fluctuation exceeding $3\sigma$ in a single independent beam is 0.13\%. Given the number of synthesized beams searched within the localization region, the total false positive probability remains below 0.4\%. 

Our deepest 15 GHz sensitivity (2.8 $\mu$Jy beam$^{-1}$ on April 9) represents $\sim 2\times$ improvement (Appendix Table \ref{tab:comparison}) over contemporary VLA D-array observations by \citet{CHIME20250316A} at 6.1–21.8 GHz (5.2–6.9 $\mu$Jy beam$^{-1}$; April 4), and is comparable to their subsequent VLA C-array observation at 9.9 GHz (2.2 $\mu$Jy beam$^{-1}$; May 30). These constraints are consistent with the absence of a compact persistent radio source, with additional evidence from observations with the European VLBI Network (EVN), High Sensitivity Array (HSA), and uGMRT \citep{CHIME20250316A}.

At the host distance of $\sim$40 Mpc, our deepest limit (15 GHz, April 9) corresponds to $\nu L_\nu < 2.4 \times 10^{35}$ erg s$^{-1}$ assuming a flat spectrum ($\alpha = 0$, with $S_\nu \propto \nu^\alpha$). 
The inferred radio luminosity is $\sim$3--4 orders of magnitude fainter than the luminosities of the three known PRSs associated with repeating FRBs ($\sim 10^{38\text{--}39}$ erg s$^{-1}$; \citealt{Chatterjee2017, Niu2022,  Bruni2025}). Combining with the Chandra 0.5--10 keV limit of $7.6 \times 10^{-15}$ erg cm$^{-2}$ s$^{-1}$ \citep{LiY2025} yields a broadband energy efficiency $\eta_{\rm radio} \lesssim 2.6 \times 10^{-4}$. Here, $\eta_{\rm radio}$ is defined as the ratio of radio luminosity to X-ray luminosity integrated over the 0.5-10 keV band. 

Image differencing between April 5 and 9 Ku-band observations revealed no residuals $>3\sigma$, ruling out transient emission or variability on 4-day timescales. While our 4-day baseline constrains short-term variability, we acknowledge that some PRSs exhibit variability on longer timescales (months to years; \citealt{Yang2024}). Future monitoring campaigns spanning weeks to months would be valuable to constrain potential long-term evolution or episodic emission.

\section{Discussion}

This non-detection of a PRS in FRB 20250316A aligns with systematic searches finding no PRSs for most one-off FRBs \citep{Aggarwal2021, Law2022}, but stands out due to FRB 20250316A's proximity and our unprecedented sensitivity. Below, we interpret these constraints through pulsar wind nebula models and empirical correlations, while highlighting implications for FRB diversity.

\subsection{Constraints from Pulsar Wind Nebula Models}

\citet{Dai2017} proposed a mechanism for persistent radio sources in which a pulsar, without surrounding supernova ejecta, sweeps its ultra-relativistic wind into ambient environments, giving rise to a non-relativistic pulsar wind nebula (PWN). The synchrotron emission from the PWN reverse shock region produces a peak flux density:
\begin{eqnarray}
    F_{\nu,\mathrm{max}} &=& 8.8 \times 10^7 \epsilon_e \epsilon_B^{11/10} \gamma_{\mathrm{min}}^{-(p-1)} \nonumber \\
    &\times& L_{w,41}^{36/25} n_{0,2}^{33/50} t_3^{18/25} \, \mu\mathrm{Jy},
\end{eqnarray}
where $L_w$ is the pulsar wind luminosity, $n_0$ is the ambient number density, $t$ is the PWN age, and we adopt $D_L = 40$ Mpc. This equation assumes an electron spectral index $p = 1.4$, tuned to FRB 20121102A, and that 15 GHz lies in the regime $\nu_a \leq \nu < \nu_c$ where $F_\nu \propto F_{\nu,\mathrm{max}} (\nu/\nu_m)^{-(p-1)/2}$. $F_\nu$ can be calculated using Equation 18 in \citet{Dai2017}.

This model assumes efficient energy transfer in dense environments ($n_0 \geq 0.1$ cm$^{-3}$); for the sparse conditions we constrain ($n_0 < 10^{-2}$ cm$^{-3}$), adiabatic expansion losses likely dominate over radiative cooling, potentially invalidating these predictions. We adopt $\epsilon_e = 0.1$ and explore magnetic field energy fractions $\epsilon_B = 10^{-2}$, $10^{-3}$, and $10^{-4}$ based on the range predicted by particle-in-cell simulations of relativistic shocks \citep[e.g.][]{Sironi2011}.

\begin{figure*}
	\begin{center}
			\includegraphics[width=0.48\textwidth]{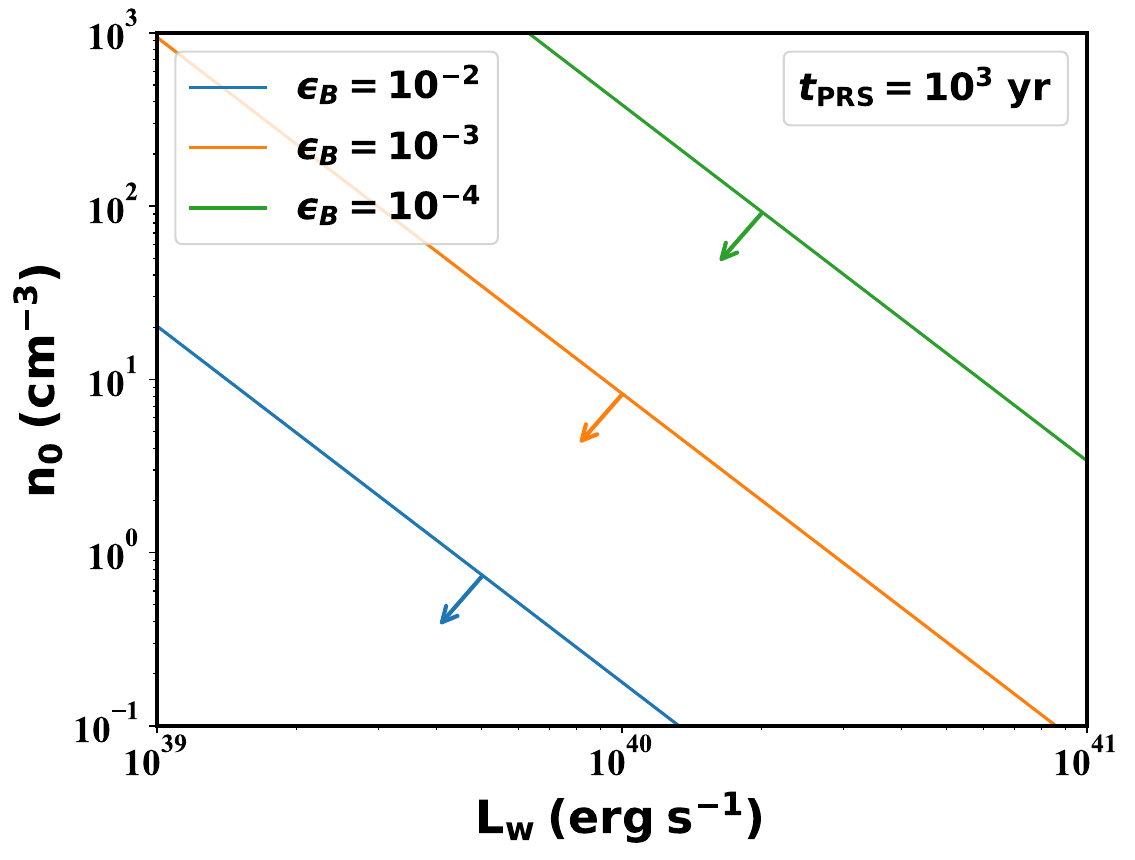}
			\includegraphics[width=0.48\textwidth]{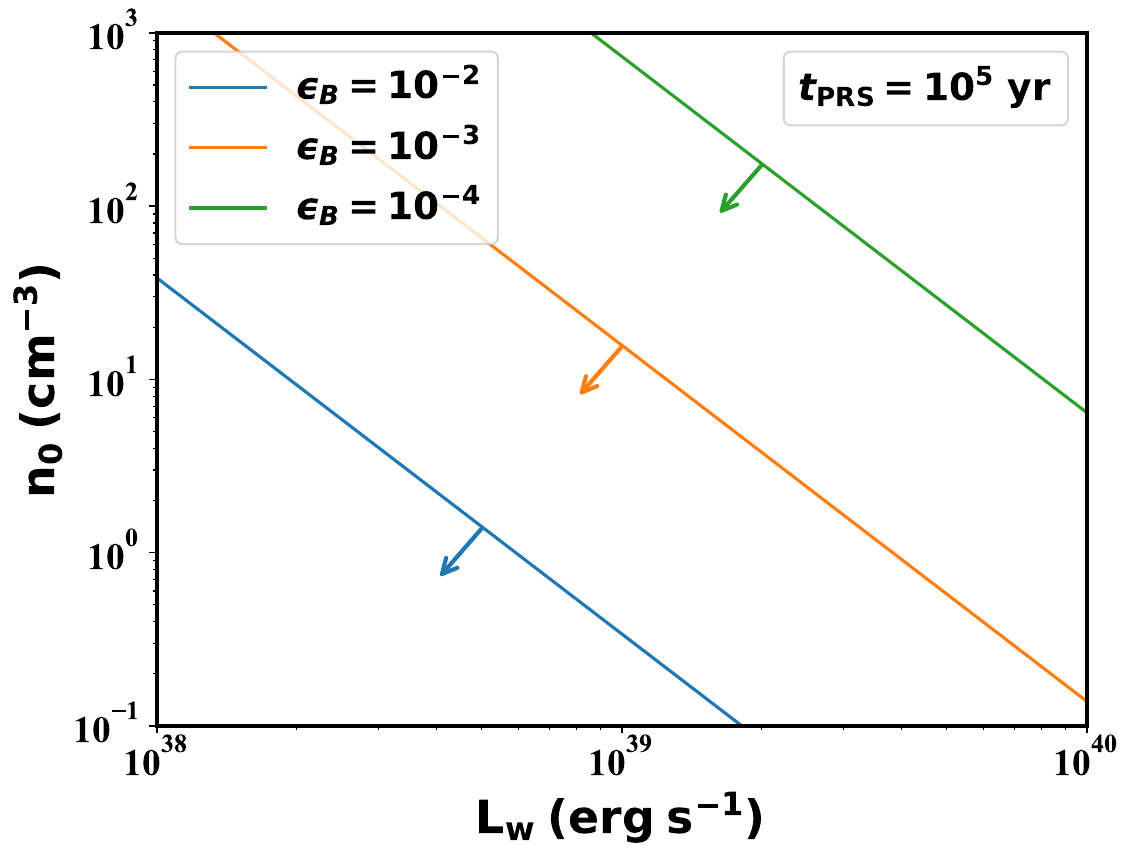}
		\caption{Parameter constraints from the observed 8.4 $\mu$Jy upper limit at 15 GHz ($3\sigma$) in the $n_0-L_w$ parameter space, based on the PRS model in \cite{Dai2017}. The green, orange, and blue solid lines correspond to $\epsilon_B=10^{-2}$, $10^{-3}$, and $10^{-4}$, respectively. Arrow symbols indicate the allowed region of parameter space. The left and right panels correspond to $t_\mathrm{PRS}=10^3\text{ yr}$ and $10^5\text{ yr}$, respectively.}
		\label{n0lw}
	\end{center}
\end{figure*}

Using our 8.4 $\mu$Jy upper limit at 15 GHz, Figure \ref{n0lw} shows the excluded parameter space in the $n_0$--$L_w$ plane for PWN ages of $10^3$ and $10^5$ years, assuming a flat spectrum ($\alpha \approx 0$). For steeper spectra ($\alpha \approx -0.7$, typical of synchrotron sources), limits tighten by factors of $\sim 2$ at lower frequencies. The curves represent detection thresholds: any combination of ambient density $n_0$ and spin-down luminosity $L_w$ above these lines would produce persistent radio emission exceeding our upper limit and thus is ruled out. Our constraints are most restrictive for young, energetic systems ($t_{\rm PRS} \leq 10^3$ yr, left panel) but remain significant for evolved scenarios ($t_{\rm PRS} \leq 10^5$ yr, right panel) . The allowed parameter space (below the curves) favors low-density environments, reduced spin-down power, or aged PWNe whose emission has faded below detectability, supporting scenarios involving evolved progenitors in sparse galactic environments rather than young magnetars in dense star-forming regions. 
Compared to FRB 20121102A \citep{Dai2017}, FRB 20250316A's proximity and our stringent constraint suggest either lower ambient density (consistent with \citealt{CHIME20250316A}) or weaker central engine activity. 
For instance, for $t = 10^5$ yr and $\epsilon_B = 10^{-3}$, our constraints require either $n_0 < 0.1$ cm$^{-3}$ for energetic engines ($L_w \sim 10^{40}$ erg s$^{-1}$) or significantly reduced spin-down power for typical ISM densities. 

We note several model limitations that warrant caution. First, the assumption of $p = 1.4$ universality is tuned to FRB 121102A. For a steeper electron spectral index $p = 2.5$ (similar to gamma-ray burst afterglows), the peak flux density becomes:
\begin{eqnarray}
    F_\mathrm{\nu,max}=&&1.21\times10^9\epsilon_e\epsilon_B^{1/2}\frac{p-2}{p-1}\gamma_\mathrm{min}^{-1}\nonumber\\
    &&\times L_{w,41}^{6/5}n_{0,2}^{3/10}t_2^{3/5}\ \mu\mathrm{Jy}.
\end{eqnarray}
This steeper index yields weaker predicted fluxes at high frequencies, relaxing $L_w$ constraints by $\sim 30\%$ for fixed $n_0$, but still favoring sparse environments.
Second, neglecting supernova ejecta could suppress early emission in young magnetar scenarios \citep[e.g.][]{Murase2016, Margalit2018}. Third, alternative ``hypernebula'' models \citep{Sridhar2022} that predict fainter, evolving emission still challenged by our limits. Recent analyses of PRS variability in FRB 20121102A and FRB 20190520B suggest dynamic environments \citep{Yang2024, Bruni2024}, motivating future monitoring.

\begin{figure}
	\begin{center}
			\includegraphics[width=0.45\textwidth]{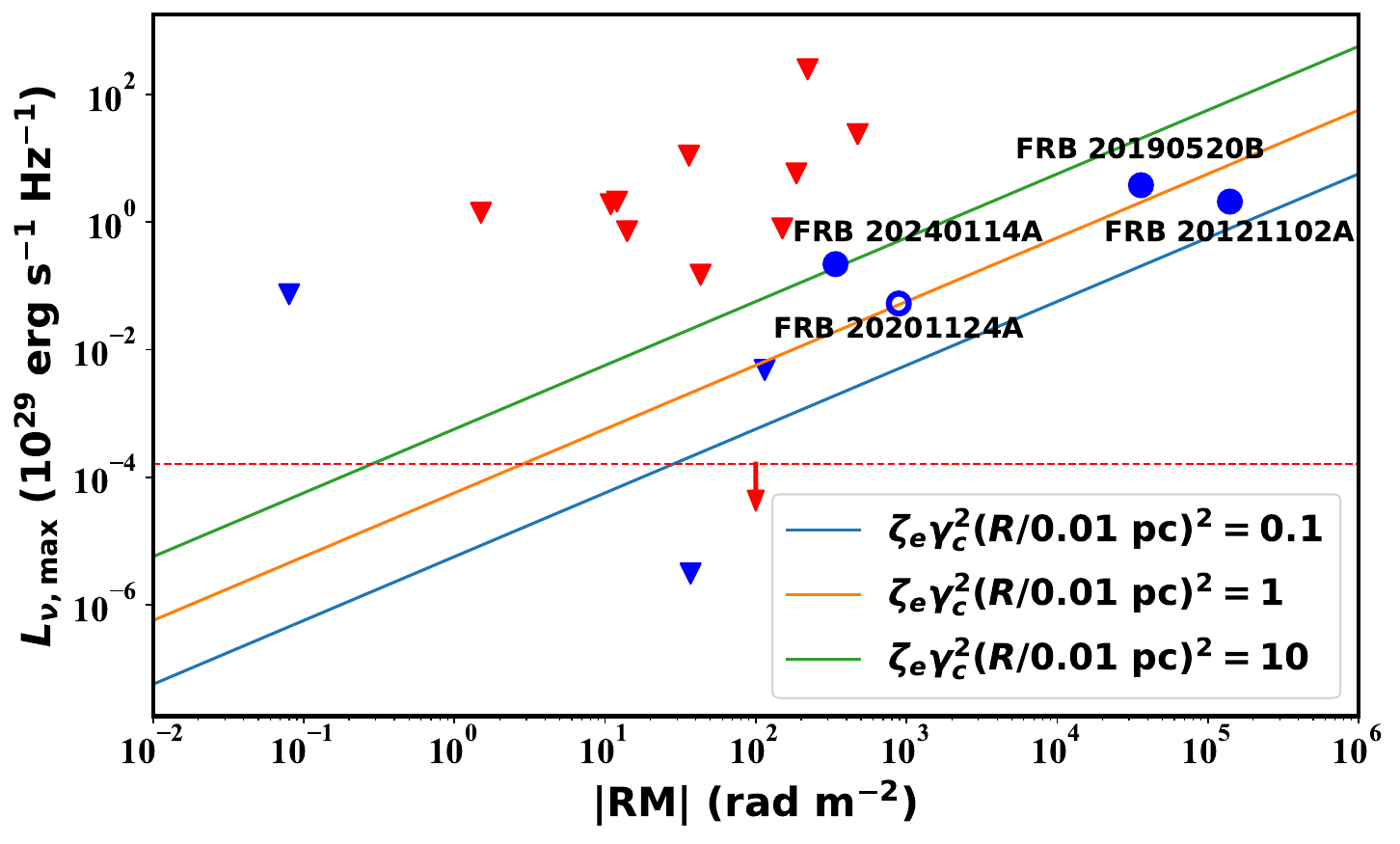}
		\caption{The observed radio flux upper limit in the PRS radio luminosity$-$FRB RM parameter space. The blue, orange, and green lines represent the cases when $\zeta_e\gamma_c^2\left(R/0.01\text{ pc}\right)^2=0.1,\text{ }1,\text{ and }10$, respectively. The $9\text{ }\mu\text{Jy}$ upper limit is marked by a red dashed line. Red and blue triangles represent the upper limit specific radio luminosities of one-off and repeating FRBs, respectively. Blue solid dots represent three FRBs \citep{Bruni2024} with measured PRS  radio luminosities (blue hollow dot represents PRS candidate for FRB 20201124A)}. Horizontal red dotted line represents the radio luminosity upper limit of $1.6 \times 10^{25}$ erg~s$^{-1}$~Hz$^{-1}$.  $\zeta_e$, $\gamma_c$, and $R$ are the fraction of synchrotron electrons, the characteristic Lorentz factor, and the radius of the plasma screen. 
		\label{lnurm}
	\end{center}
\end{figure}

\subsection{Speculative Extrapolation of PRS--RM Relation}

An empirical correlation between PRS luminosity and FRB rotation measure (RM) has been established for repeating FRBs with detected persistent emission \citep{Yang2020, Yang2022, Bruni2024}. This relation suggests that brighter PRSs are associated with higher RMs, potentially reflecting stronger magnetic fields or denser plasma environments. Assuming the empirical PRS–RM relation extends to one-off FRBs, despite remaining highly speculative, our $<9$ $\mu$Jy upper limit intersects the three model curves at $\vert{\rm RM}\vert \lesssim$  0.3, 3, and 30 rad~m$^{-2}$, respectively. This range covers the measured value of $+16.79\pm0.85$ rad~m$^{-2}$ reported by \citet{CHIME20250316A}. Taking the estimated host galaxy DM contribution of $\sim 60$ pc cm$^{-3}$ together with the $\vert{\rm RM}\vert$ yields a mean line-of-sight magnetic field strength of $\lesssim$ a few $\mu$G. Such a weak, diffuse environment is incompatible with the dense, magnetized sites expected for young magnetars, which have surface dipole fileds $B \sim 10^{14}$--$10^{15}$ G \citep{Kaspi2017} and are are predicted to reside in compact, high-RM magnetar-wind/SN nebulae \citep[e.g.,][]{Michilli2018, Metzger2019, Margalit2019}, but matches expectations for older neutron-star engine expanding in a halo-like medium  \citep[e.g.,][]{Tumlinson2017, Margalit2018, Sridhar2022} and for the radio luminosities of evolved Galactic PWNe  \citep[e.g.,][]{Gaensler2006, Green2019}. 

However, extending this relation to one-off events may face limitations. The calibration sample remains small ($N = 3$) and potentially biased toward repeating FRBs that may preferentially inhabit magnetized environments enabling sustained emission. Repeaters comprise $\sim 5-10\%$ of the known FRB population and may represent a distinct physical class. Critically, one-off and repeating bursts could arise from fundamentally different mechanisms, i.e., catastrophic events like neutron star mergers versus persistent magnetar activity, invalidating the assumption of universal PRS--RM scaling across FRB types. Moreover, the PRS-RM correlation may be driven by selection effects, as FRBs in dense environments are more likely to be detected as repeaters due to enhanced burst rates \citep{Michilli2018}. The absence of polarimetric measurements for FRB 20250316A prevents direct testing of this speculation. We therefore emphasize this extrapolation as highly speculative, providing a tentative environmental constraint that awaits validation through direct polarimetric measurements of FRB 20250316A or expanded statistical samples of FRB--PRS associations.

\subsection{Implications for FRB Diversity}

Our upper limit of $< 8.4 \mu$Jy at 15 GHz, combined with the comprehensive analysis by \citet{CHIME20250316A}, represents one of the deepest searches for a PRS associated with a one-off FRB. For comparison, \citet{Law2022} achieved typical limits of 15-30 $\mu$Jy for a sample of 7 one-off FRBs at comparable frequencies, while \citet{Aggarwal2021} reported limits of 10-50 $\mu$Jy for 10 FRBs.  The combination of FRB 20250316A's proximity and our deep observations provides constraints $\sim$3-5 times more stringent than typical searches, yet still yields no detection. This strengthens the emerging dichotomy between repeating FRBs with luminous PRSs and the broader one-off population.

Our luminosity upper limit ($\nu L_\nu < 2.4 \times 10^{35}$ erg s$^{-1}$) provides crucial context when compared to Galactic pulsar wind nebulae. This constraint aligns remarkably well with evolved PWNe around aged pulsars: the Vela PWN exhibits $\sim 10^{34}$--$10^{35}$ erg s$^{-1}$ at radio frequencies \citep{Green2019}, while young, energetic systems like the Crab Nebula radiate $\sim 10^{36}$--$10^{37}$ erg s$^{-1}$. Our limit falls orders of magnitude below the nebular luminosities expected from young magnetars ($\sim 10^{37}$--$10^{38}$ erg s$^{-1}$), suggesting any progenitor resembles an evolved neutron star rather than a recently born, highly magnetized object.

This comparison strengthens the emerging paradigm of FRB progenitor diversity. The sharp dichotomy between repeating FRBs with luminous persistent sources and one-off events without detectable nebular emission suggests fundamentally different physical mechanisms rather than a simple evolutionary sequence. Repeaters, associated with star-forming regions and young stellar populations \citep{Heintz2020, Tendulkar2021}, may arise from young magnetars in dense environments capable of sustaining luminous nebulae. In contrast, one-offs like FRB 20250316A reside in clean, low-density, weak magnetic field environments, favoring cataclysmic origins such as neutron star mergers or giant flares from old magnetars in sparse galactic halos. The offset of $190\pm20$ pc from the nearest star-formation region \citep{CHIME20250316A} is similar to that of FRB 20180916B ($\sim 250$ pc), suggesting possible evolutionary scenarios for aged neutron stars.

Alternatively, the apparent diversity could reflect observational selection effects rather than intrinsic physical differences. Enhanced burst activity and precise localization may preferentially enable PRS detection around repeaters, while similar but fainter nebulae around one-offs remain below current sensitivity thresholds. Testing this hypothesis requires expanding PRS searches to larger samples of well-localized one-off FRBs.

\section{Summary}

Our deep VLA observations of FRB 20250316A yield no detection of a persistent radio source across 6--15 GHz, with stringent upper limits of $<8.4$ $\mu$Jy at 15 GHz, ruling out luminous nebulae comparable to those around repeating FRBs and any transient variability on 4-day timescales.

Combined with the source's proximity ($\sim$40 Mpc) and one-off nature, these constraints challenge models requiring young, energetic pulsar wind nebulae for all FRBs, instead favoring evolved progenitors in sparse environments or distinct cataclysmic origins such as neutron star mergers. Comparison with Galactic analogs suggests similarity to aged rather than young PWN. The empirical PRS--RM relation, if applicable beyond repeaters, would predict low RMs consistent with clean magnetoionic environments, though this extrapolation remains highly speculative given potential selection biases and the small calibration sample.

These results strengthen the emerging picture of FRB progenitor diversity, with repeaters and one-offs potentially arising from fundamentally different physical mechanisms. Our results, combined with similar non-detections for other one-off FRBs, suggest that bright persistent emission may not be a universal feature of FRB progenitors. This supports a framework where repeating and non-repeating FRBs arise from distinct physical mechanisms or evolutionary stages, though expanded samples and multi-wavelength campaigns are essential to confirm this paradigm. Future observations with next-generation facilities (\textit{SKA}, \textit{ngVLA}) and direct polarimetric measurements will be crucial for resolving whether this apparent diversity reflects true physical distinctions or observational biases.

\section{acknowledgments}
We thank the VLA DDT committee for approving and rapidly scheduling the observations (project code 25A-435). 
This work is supported by the National SKA Program of China (2020SKA0120200, 2022SKA0130103, 2020SKA0120302), National Natural Science Foundation of China (12393812, 12503018), the Strategic Priority Research Program of the Chinese Academy of Sciences (XDB0550300) and FAST Special Program (NSFC 12041301). 
T.A. acknowledges support from the Shanghai Oriental Talent Project and Xinjiang Tianchi Talent Program.
A.W. acknowledges the 18th special financial grant (2025T180874) and the 77th General Program grant (2025M773197) from the China Postdoctoral Science Foundation. 
The data processing made use of the compute resource of the China SKA Regional Centre \citep{2019NatAs...3.1030A,2022SCPMA..6529501A}. 
The National Radio Astronomy Observatory is a facility of the National Science Foundation operated under cooperative agreement by Associated Universities, Inc. 

\vspace{5mm}
\facilities{VLA}  
\software{astropy \citep{2013A&A...558A..33A,2018AJ....156..123A}, CASA \citep{McMullin2007} }

\bibliography{frb250316a}{}
\bibliographystyle{aasjournal}

\appendix   
\section{Detailed VLA Data Processing} \label{app:reduction}

\subsection{Observational Setup and Strategy}

We conducted VLA observations (project code: 25A-435, PI: T. An) of FRB 20250316A over two epochs: 2025 April 5 00:59:33--02:47:25 and 2025 April 9 01:08:08--02:57:10.
The VLA pointing was centered on the CHIME/Outriggers VLBI localization of FRB~20250316A \mbox{(R.A.\ 12$^{\rm h}$09$^{\rm m}$44\fs319, Dec.\ +58\degr50\arcmin56\farcs708} (J2000). Primary-beam attenuation at that offset is $<1\%$ and negligible for our sensitivity goals. VLA was in the D-configuration during the observations, delivering synthesized beams of $\sim4''$--14$''$ over our frequencies. Integration times were set to reach the thermal-noise floor of $\sim$3--6~$\mu$Jy~beam$^{-1}$, sufficient to detect potential PRSs comparable to the faintest known examples.

We observed in C (4--8~GHz), X (8--12~GHz), and Ku (12--18~GHz) bands. The initial C+X+Ku coverage on April~5 was designed to enable spectral index determination if a PRS was detected, while \emph{Ku band (15~GHz)} was prioritized as the primary detection band owing to its higher resolution (few-arcsec beams in D array),  sufficient to disentangle a compact PRS from host-galaxy star-forming clumps. Frequencies below $\sim$4~GHz were \emph{not} scheduled, as extended star-formation complexes near the FRB position dominate at low frequencies (see the 3.2-GHz VLA image in \citealt{CHIME20250316A}) and would limit dynamic range and confuse compact-source searches. Observations used short phase-referencing cycles (7--10~min on-source bracketed by $\sim$1~min on a nearby calibrator) to track atmospheric phase variations. Scans were constrained to elevations $>30^\circ$ to minimise airmass-dependent systematics.

\subsection{Calibration Strategy and Imaging}

Primary calibrator 3C~286 was observed for 5 minutes at the beginning and end of each session. We adopted the \cite{2017ApJS..230....7P} flux density scale, providing absolute flux accuracy of $\sim$2\% at our frequencies. Bandpass solutions derived from 3C~286 correct for instrumental spectral response variations across the 4--6 GHz bandwidths per band. We used  \texttt{CASA} version 6.4.1 for calibration and imaging.

The absolute flux density scale was verified by comparing derived 3C~286 flux densities with contemporary VLA monitoring data, confirming consistency within 1.5\% across all bands. Secondary calibrators J1219+4829 (C- and X-band) and J1152+4939 (Ku-band) were chosen from the VLA calibrator database based on: (1) angular proximity (10.4$^\circ$ from target), (2) flux density stability documented in VLA monitoring archives spanning 2020--2025, and (3) compact structure verified by VLA or Very Long Baseline Array (VLBA).

Complex gain solutions were derived using 1-minute solution intervals, sufficient to track tropospheric phase variations while maintaining adequate signal-to-noise for stable amplitude solutions. Gain solution quality metrics show phase rms $<15^\circ$ and amplitude scatter $<3\%$ across all antennas and frequencies.

\begin{figure*}
\centering
\includegraphics[width=0.48\textwidth]{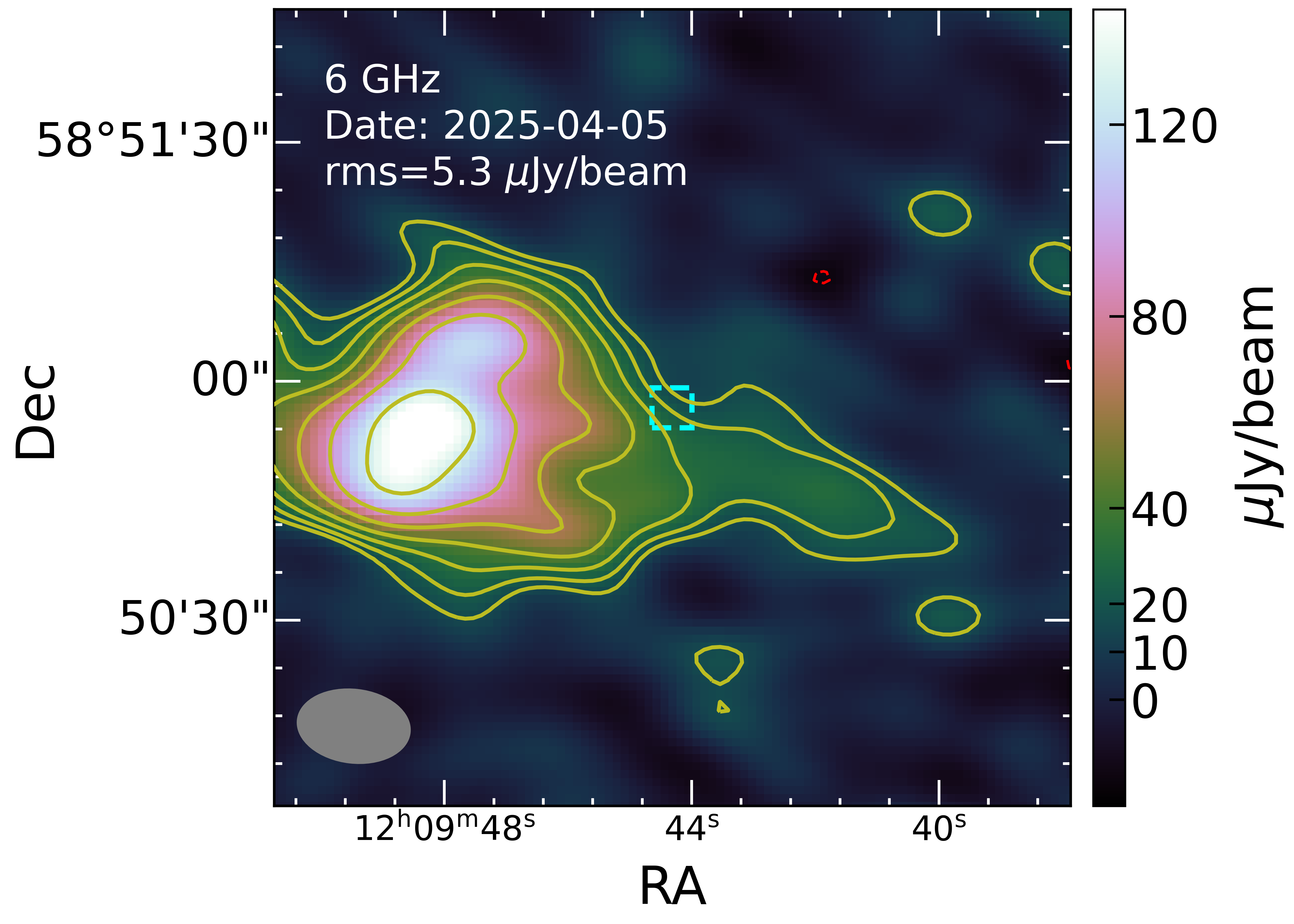}
\includegraphics[width=0.48\textwidth]{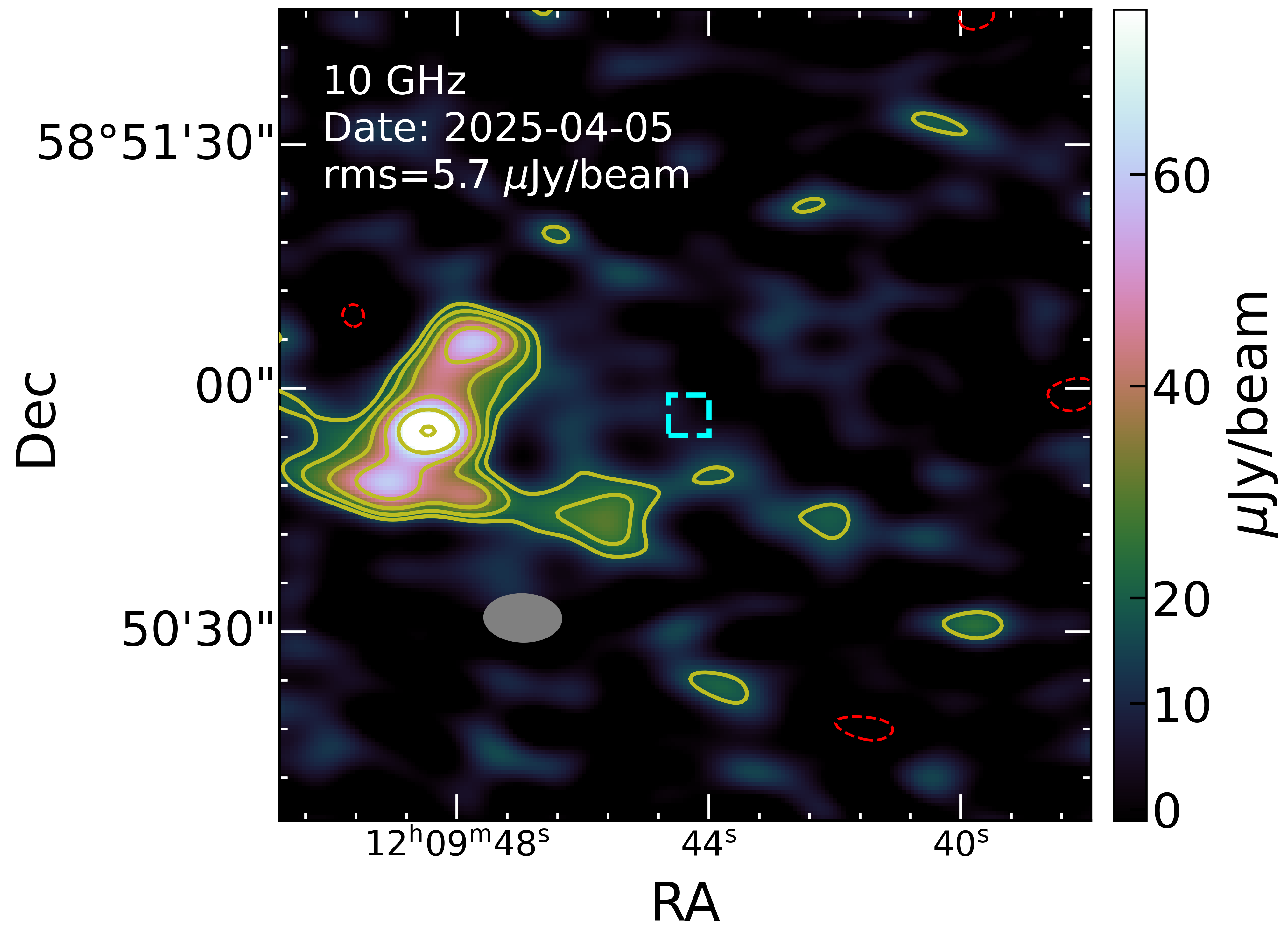}
\includegraphics[width=0.48\textwidth]{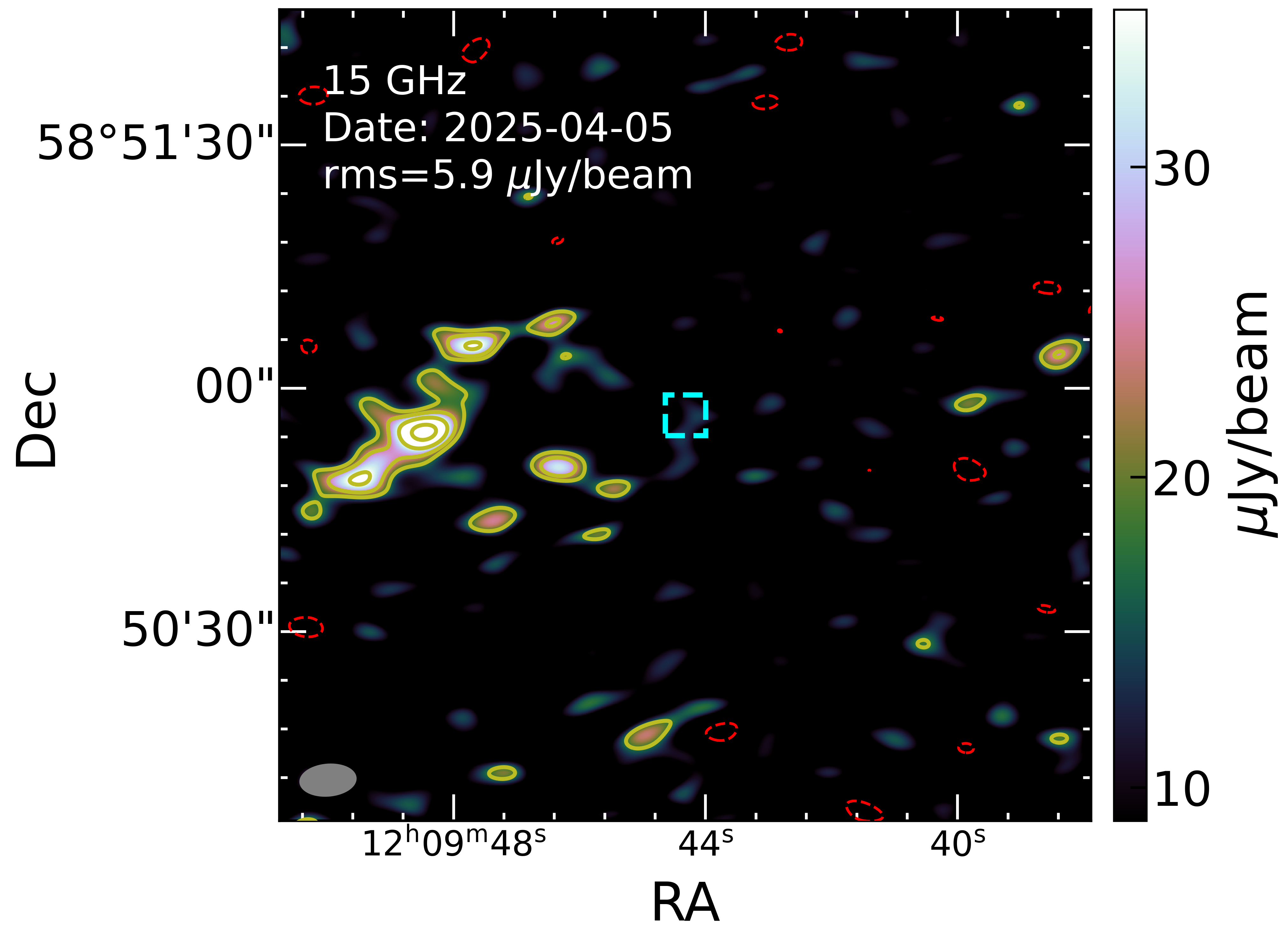}
\includegraphics[width=0.48\textwidth]{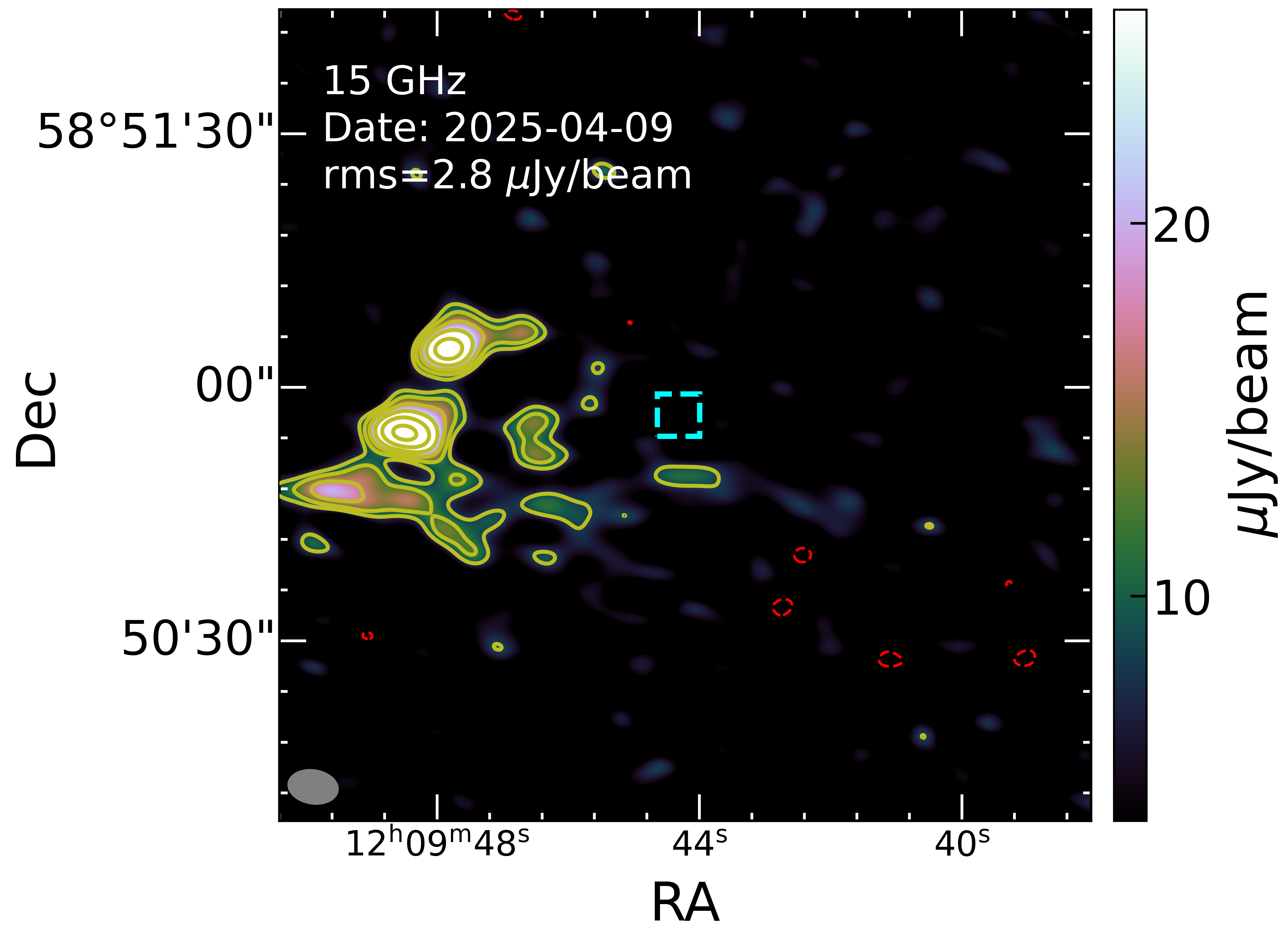}
\caption{VLA images demonstrating the non-detection of persistent radio emission at the FRB 20250316A position across different bands and epochs. 
\textit{Top left:} C-band (6 GHz, 2025 April 5) with synthesized beam 14$''$ $\times$ 9$''$ and rms noise 5.3 $\mu$Jy beam$^{-1}$. 
\textit{Top right:} X-band (10 GHz, 2025 April 5) with synthesized beam 9$''$ $\times$ 6$''$ and rms noise 5.7 $\mu$Jy beam$^{-1}$. 
\textit{Bottom left:} Ku-band (15 GHz, 2025 April 5) with synthesized beam 7$''$ $\times$ 4$''$ and rms noise 5.9 $\mu$Jy beam$^{-1}$. 
\textit{Bottom right:} Ku-band (15 GHz, 2025 April 9) with synthesized beam 6$''$ $\times$ 4$''$ and rms noise 2.8 $\mu$Jy beam$^{-1}$. 
The bright-blue dashed square, centered at the FRB position (RA 12h09m44.319s, Dec +58$^\circ$50$'$56.708$''$), has a size of $5''$ on each side, while the gray ellipse in the bottom-left corner of each panel indicates the synthesized beam.
Contour levels are drawn at $3\sigma$ and increased by factors of $\sqrt{2}$, with the red dashed contour marking $-3\sigma$. 
No significant emission ($>3\sigma$) is detected at the FRB position in any band or epoch. }
\label{fig:nondetection}
\end{figure*}

Final images were produced with multi-frequency synthesis (MFS) combining all spectral windows per band. We adopted two Taylor terms (\texttt{nterms=2}) to capture in-band spectral indices. To balance resolution and point-source sensitivity we used Briggs weighting with \texttt{robust=0.5}. Deconvolution employed multi-scale CLEAN with scales $[0,8,20]$ pixels to recover compact to moderately extended emission. CLEAN masks were restricted to a $30^{\prime\prime}$ radius around the FRB position to avoid cleaning noise. Imaging was stopped when the peak residual fell below three times the thermal noise estimate.
Figure \ref{fig:nondetection} shows the non-detection of persistent radio emission at the FRB 20250316A position across different bands and epochs.

\subsection{Comparison with Contemporary VLA Observations}
\label{app:comparison}

To place our sensitivities in context, Table~\ref{tab:comparison} summarizes our VLA D-array results alongside contemporary VLA follow-ups. Our 2025-04-05 D-array images reach rms levels comparable to the results reported by \citet{CHIME20250316A} (differences $< 10\%$ at similar frequencies). The dedicated Ku-band session on 2025-04-09 attains our deepest image (rms $=2.8~\mu$Jy~beam$^{-1}$ at 15~GHz), a factor of $\sim 2$ better than the D-array results from 2025-04-05, owing to focused time in a single high-frequency band and excellent weather. The later 2025-05-30 9.9~GHz C-array image from CHIME/FRB reaches $2.2~\mu$Jy~beam$^{-1}$, reflecting the expected trade-off: higher angular resolution mitigates confusion and can yield lower rms on compact sources, whereas D-array offers superior surface-brightness sensitivity to extended emission; C-array has lower confusion limits than D-array, thus achieves a lower noise floor for the same integration time. 
Thermal-noise expectations were evaluated with the standard interferometric radiometer equation,
\[
\sigma \;\approx\; \frac{\mathrm{SEFD}}{\eta_s\,\sqrt{N_{\rm pol}\,N_{\rm ant}(N_{\rm ant}-1)\,\Delta\nu\,\Delta t}}\;,
\]
where $\mathrm{SEFD}$ is the system-equivalent flux density (set by $T_{\rm sys}$ and aperture efficiency), $\eta_s$ is the system/imaging efficiency, $N_{\rm pol}=2$, $\Delta\nu$ is the effective bandwidth, and $\Delta t$ is the on-source integration. Using measured bandwidths of 4–6~GHz and on-source times of $18$–$76$~min per band, the predicted image rms is $3.1$–$5.1~\mu$Jy~beam$^{-1}$, in good agreement with the observed $2.8$–$5.9~\mu$Jy~beam$^{-1}$ range.

Collectively, these multi-epoch, multi-configuration observations provide robust constraints spanning 4--18 GHz with sensitivities ranging from 2.2--7.2 $\mu$Jy beam$^{-1}$. The consistency of non-detections across different observing epochs, array configurations, and processing strategies strengthens confidence in the absence of persistent radio emission associated with FRB 20250316A at the probed sensitivity levels.

\begin{table}
\centering
\caption{Comparison of VLA Observations for FRB 20250316A}
\label{tab:comparison}
\begin{tabular}{lccccc}
\hline
Date & Array & Frequency  & RMS Noise & Reference \\
 & Config. & (GHz)  & ($\mu$Jy beam$^{-1}$) & \\
\hline
2025-04-04 & D & 3.2 &13.6 & \citet{CHIME20250316A} \\
2025-04-04 & D & 6.1 & 5.6 & \citet{CHIME20250316A} \\
2025-04-04 & D & 9.9 & 5.2 & \citet{CHIME20250316A} \\
2025-04-04 & D &21.8 & 6.9 & \citet{CHIME20250316A} \\
2025-04-05 & D & 6.0 & 5.3 & This work \\
2025-04-05 & D & 10.0& 5.7 & This work \\
2025-04-05 & D & 15.0& 5.9 & This work \\
2025-04-09 & D & 15.0& 2.8 & This work \\
2025-05-10 & D & 3.2 &13.5 & \citet{CHIME20250316A} \\
2025-05-10 & D & 6.1 & 5.8 & \citet{CHIME20250316A} \\
2025-05-10 & D & 9.9 & 5.6 & \citet{CHIME20250316A} \\
2025-05-10 & D &21.8 & 7.8 & \citet{CHIME20250316A} \\
2025-05-30 & C & 9.9 & 2.2 & \citet{CHIME20250316A} \\
\hline
\end{tabular} 
\end{table}

\end{document}